 \documentclass[final,5p,times,twocolumn]{elsarticle}

\usepackage{amssymb}
\usepackage{lipsum}
\usepackage{braket}
\usepackage{lineno}
\usepackage{hyperref}
\usepackage{amsmath} 
\usepackage{xcolor}

\journal{Physics Letters B}

\begin{document}

\begin{frontmatter}



\title{Bosonic Spin-1 SOPHY}


\author[a]{Armando De la C. Rangel-Pantoja\corref{cor1}}
\cortext[cor1]{Corresponding author}
\ead{adlc.rangelpantoja@ugto.mx}
\author[b]{I. D\'iaz-Salda\~na}
\author[a,c,d]{Carlos A. Vaquera-Araujo}

\affiliation[a]{organization={Departamento de F\'isica, Divisi\'on de Ciencias e Ingenier\'ias, Campus Le\'on, Universidad de
  Guanajuato},
            addressline={ Loma del Bosque 103, Lomas del Campestre}, 
            city={Le\'on},
            postcode={37150}, 
            state={Guanajuato},
            country={Mexico}}
 \affiliation[b]{organization={Unidad Acad\'emica de F\'isica, Universidad Aut\'onoma de Zacatecas},
             addressline={Calzada Solidaridad esquina con Paseo la Bufa S/N},
             city={Zacatecas},  postcode={98060},
             state={Zacatecas},
             country={Mexico}}            
\affiliation[c]{organization={Secretar\'ia de Ciencia, Humanidades, Tecnolog\'ia e Innovaci\'on},
             addressline={Insurgentes Sur 1582. Colonia Cr\'edito Constructor},
             city={Benito Ju\'arez},
             postcode={03940},
             state={Ciudad de M\'exico},
             country={Mexico}}
 \affiliation[d]{organization={Dual CP Institute of High Energy Physics},
             city={Colima},
            postcode={28045},
            state={Colima},
             country={Mexico}}

\begin{abstract}
In this work we study the canonical quantization of a second-order pseudo-Hermitian field theory for massive spin-1 bosons transforming under the $(1,0)\oplus(0,1)$ representation of the restricted Lorentz Group and satisfying the Klein-Gordon equation.
\end{abstract}




\end{frontmatter}




\section{Introduction}
\label{intro}
In recent years, there has been renewed interest in the study of pseudo-Hermitian quantum field theories, a topic pioneered by the study of symplectic fermions in~\cite{LeClair:2007iy}, where scalar fields that satisfy the Klein-Gordon (KG) equation evade the usual spin-statistics connection by relaxing the hermiticity requirement in favor of a less stringent one. The work in~\cite{LeClair:2007iy} was the first attempt to build a pseudo-Hermitian quantum field theory based on the success of pseudo-Hermitian quantum mechanics~\cite{Mostafazadeh:2001jk,Mostafazadeh:2008pw}, where a suitable redefinition of the inner product by means of an auxiliary operator guarantees that the energy eigenvalues are real and that the time evolution is unitary, generalizing the results of $PT$-symmetric quantum mechanics~\cite{Bender:1998ke}, a well-established and fertile discipline with applications in optics, Bose-Einstein condensates, metamaterials, electric circuits, acoustics and mechanical systems \cite{Christodoulides:2018arz}.
Since then, several pseudo-Hermitian quantum field theories have been formulated. Notably, for some of them the dynamics are dictated by the KG equation and therefore can be classified as second-order pseudo-Hermitian theories, including massive spin-$1/2$ fermions~\cite{Ferro-Hernandez:2023ymz}, spin-$1/2$ Elko fermions and bosons with two-fold Wigner degeneracy~\cite{Ahluwalia:2023slc} (for a recent formulation of a Hermitian formalism for Elko fields see \cite{deGracia:2024umr,deGracia:2025omq}), and a simple theory for massive spin-$1/2$ bosons, built directly from Dirac spinors~\cite{Rangel-Pantoja:2025mhg}. The case of a first-order pseudo-Hermitian theory has also been studied in~\cite{Ahluwalia:2022zrm}, where a spin-$1/2$ field satisfying the Dirac equation behaves as a boson. More recently, a new spin-statistics theorem for a class of free pseudo-Hermitian quantum field theories was established in \cite{Bai:2025pvb}, unlocking the formulation of new theories with flipped statistics, such as those studied in \cite{LeClair:2007iy,Rangel-Pantoja:2025mhg,Ahluwalia:2022zrm}.

In this paper, we study the first Second-Order Pseudo-Hermitian theorY (SOPHY) for massive spin-1 bosons transforming in the $(1,0)\oplus(0,1)$ representation of the Restricted Lorentz Group (RLG) $SO(1,3)^+$, which consists of proper orthocronous Lorentz transformations.  The main motivation to study this theory resides on the fact that it admits a rich set of renormalizable self-interactions, and consequently it is the only known renormalizable theory for massive spin-$1$ particles that does not involve additional scalar fields, as in the Higgs \cite{Englert:1964et,Higgs:1964pj,Guralnik:1964eu} and Stueckelberg mechanisms \cite{Ruegg:2003ps}. Besides, the fields described by this class of theories have mass dimension one by construction, and if they transform as singlets under the gauge symmetries of the Standard Model (SM), they cannot decay into SM fields via renormalizable operators, since they always come into pairs in interactions. Thus, the least massive pseudo-Hermitian particle can play the role of a weakly interacting massive
particle (WIMP) dark matter candidate.   

The paper is organized as follows: In Section~\ref{sec:fields}, The most general field in the $(1,0)\oplus(0,1)$ representation that satisfies the KG equation is built from first principles; in Section~\ref{sec:quant}, the Canonical Quantization of the theory is presented, showing that the resulting SOPHY is causal, the Hamiltonian is bounded from below with real eigenvalues, the global symmetries of the theory are identified, and its invariance under the discrete symmetries of Charge conjugation, Parity and Time-reversal is established; finally, in Section~\ref{sec:conc} we draw some conclusions about our results.

\section{Fields in the \texorpdfstring{$(1,0)\oplus(0,1)$}{(1,0)+(0,1)} representation of the RLG}
\label{sec:fields}
In order to make this paper as self-contained as possible, we build the fields from scratch, following the lines of~\cite{Napsuciale:2022usn} but adapting the notation to resemble that of~\cite{Langacker:2017uah}.
\subsection{The $(1,0)$ and $(0,1)$ representations}
The generators of the $SO(1,3)^+$ group satisfy the commutation relations
\begin{gather}
[J^i,J^j] = i\epsilon^{ijk}J^k, \,\, [J^i,K^j] = i\epsilon^{ijk}K^k,\,\,
[K^i,K^j] = -i\epsilon^{ijk}J^k.
\end{gather}
This algebra can be split into two independent $\mathfrak{su}(2)$ algebras by defining the operators
\begin{equation}
A^i=\frac{1}{2}(J^i+i K^i),\quad B^i=\frac{1}{2}(J^i-i K^i),
\end{equation}
that satisfy
\begin{gather}
[A^i,A^j] = i\epsilon^{ijk}A^k, \quad [B^i,B^j] = i\epsilon^{ijk}B^k,\quad
[A^i,B^j] = 0.
\end{gather}
Thus, the $SO(1,3)^+$ irreducible representations (irreps) are characterized by the eigenvalues of the Casimir operators $\mathbf{A}^2$, $\mathbf{B}^2$, labeled by $(a,b)$, respectively. 
The corresponding group elements for a column vector transforming under the $(a,b)$ irrep become, in this basis
\begin{equation}
S(\boldsymbol{\theta},\boldsymbol{\varphi})=e^{-i(\mathbf{J}\cdot\boldsymbol{\theta}+\mathbf{K}\cdot\boldsymbol{\varphi})}=e^{-i(\mathbf{A}+\mathbf{B})\cdot\boldsymbol{\theta}-(\mathbf{A}-\mathbf{B})\cdot\boldsymbol{\varphi}},
\end{equation}
with rotation angles $\boldsymbol{\theta}=\theta\mathbf{\hat{m}}$ and rapidities $\boldsymbol{\varphi}=\varphi\mathbf{\hat{n}}$.
In particular, for the \emph{left-handed} $(1,0)$ irrep, with $\mathbf{B}=0$, the Lorentz transformation takes the form
\begin{gather}
S_L(\boldsymbol{\theta},\boldsymbol{\varphi})=e^{-i\mathbf{J}^{(1)}\cdot\boldsymbol{\theta}-\mathbf{J}^{(1)}\cdot\boldsymbol{\varphi}}\equiv R(\boldsymbol{\theta})B_L(\boldsymbol{\varphi}),
\end{gather}
while for the \emph{right-handed} $(0,1)$ irrep, with $\mathbf{A}=0$, we have
\begin{gather}
S_R(\boldsymbol{\theta},\boldsymbol{\varphi})=e^{-i\mathbf{J}^{(1)}\cdot\boldsymbol{\theta}+\mathbf{J}^{(1)}\cdot\boldsymbol{\varphi}}\equiv R(\boldsymbol{\theta})B_R(\boldsymbol{\varphi}),
\end{gather}
where $\mathbf{J}^{(1)}$ are the $SU(2)$ generators for $j=1$
\begin{gather}
    J^{1\,(1)}
    =
    \begin{pmatrix}
        0 & \frac{1}{\sqrt{2}} & 0 \\
        \frac{1}{\sqrt{2}} & 0 & \frac{1}{\sqrt{2}} \\
        0 & \frac{1}{\sqrt{2}} & 0
    \end{pmatrix},
    \quad
    J^{2\,(1)}
    =
    \begin{pmatrix}
        0 & -\frac{i}{\sqrt{2}} & 0 \\
        \frac{i}{\sqrt{2}} & 0 & -\frac{i}{\sqrt{2}} \\
        0 & \frac{i}{\sqrt{2}} & 0
    \end{pmatrix},\nonumber\\
    J^{3\,(1)}
    =
    \begin{pmatrix}
        1 & 0 & 0 \\
        0 & 0 & 0 \\
        0 & 0 & -1
    \end{pmatrix}.
\end{gather}
Using the fact that these generators satisfy the following relation
\begin{equation}
[\mathbf{\hat{m}}\cdot\mathbf{J}^{(1)}]^3=\mathbf{\hat{m}}\cdot\mathbf{J}^{(1)},
\end{equation}
it can be shown that the explicit form of the rotations is
 \begin{equation}
\begin{split}
R(\boldsymbol{\theta})&=\mathbf{1}-i\sin\theta\mathbf{\hat{m}}\cdot\mathbf{J}^{(1)}+(\cos\theta-1)[\mathbf{\hat{m}}\cdot\mathbf{J}^{(1)}]^{2}.
\end{split}
\end{equation}
while under boosts, left- and right-handed fields transform according to 
 \begin{equation}
\begin{split}
B_{L,R}(\boldsymbol{\varphi})&=\mathbf{1}\mp\sinh\varphi\mathbf{\hat{n}}\cdot\mathbf{J}^{(1)}+(\cosh\varphi-1)[\mathbf{\hat{n}}\cdot\mathbf{J}^{(1)}]^{2}.
\end{split}
\end{equation}
Since $SU(2)$ is a pseudoreal group, there exists a matrix $\epsilon$ that relates the rotation matrix to its conjugate as
$
R(\boldsymbol{\theta})^*=\epsilon^{-1} R(\boldsymbol{\theta})\epsilon.
$
In our case at hand, this operator is 
\begin{align}
    \epsilon=&
    e^{-i\pi J^{2\,(1)}}
    =
    \begin{pmatrix}
        0 & 0 & 1 \\
        0 & -1 & 0 \\
        1 & 0 & 0 \\
    \end{pmatrix},
\end{align}
satisfying $\epsilon=\epsilon^{-1}=\epsilon^\dagger=\epsilon^{T}=\epsilon^{*}$, and 
\begin{gather}
    [\mathbf{\widehat{m}}\cdot \mathbf{J}^{(1)} ]^{*} 
    =-\epsilon [\mathbf{\widehat{m}}\cdot \mathbf{J}^{(1)} ] \epsilon,\quad   [\mathbf{\widehat{m}}\cdot \mathbf{J}^{(1)} ]^2{}^{*} 
    =\epsilon[\mathbf{\widehat{m}}\cdot \mathbf{J}^{(1)} ]^2\epsilon.\label{conjeps}
\end{gather}
In terms of momentum, we have $\cosh\varphi=\gamma=\frac{\omega_{\mathbf{p}}}{m}$, $\sinh\varphi=\gamma\beta=\frac{|\mathbf{p}|}{m}$, with $\omega_{\mathbf{p}}=+\sqrt{\mathbf{p^2}+m^2}$, thus the boosts are explicitly
\begin{equation}\label{BoostLR}
\begin{split}
B_{L,R}(\mathbf{p})
&=\mathbf{1}\mp\frac{\mathbf{p}\cdot\mathbf{J}^{(1)}}{m}+\frac{[\mathbf{p}\cdot\mathbf{J}^{(1)}]^{2}}{m(\omega_{\mathbf{p}}+m)},
\end{split}
\end{equation}
and satisfy $B^\dagger_{L,R}(\mathbf{p})=B_{L,R}(\mathbf{p})$.
In addition, the boosts for left, and right-handed fields are orthogonal $ B_{R,L}(\mathbf{p})B_{L,R}(\mathbf{p})=1$
and, therefore, they are their respective inverses $B_{R,L}^{-1}(\mathbf{p})
    =
    B_{L,R}(\mathbf{p}).$

The squared left and right boost operators can be written as
\begin{align}
    B_{L}(\mathbf{p})^2=&
    \tau^{\mu\nu} p_{\mu} p_{\nu}/m^2\equiv \tau(p)/m^2,
    \\
    B_{R}(\mathbf{p})^2
=&
    \bar{\tau}^{\mu\nu} p_{\mu} p_{\nu}/m^2\equiv \bar{\tau}(p)/m^2,
\end{align}
with
\begin{gather}
    \tau^{00}=1, \quad \tau^{0i}=\tau^{i0}=J^{i\,(1)},\quad\tau^{ij}=g^{ij}\mathbf{1}+\{J^{i\,(1)},J^{j\,(1)}\},
   \\
  \bar{\tau}^{00}=1, \quad \bar{\tau}^{0i}=\bar{\tau}^{i0}=-J^{i\,(1)},\quad\bar{\tau}^{ij}=g^{ij}\mathbf{1}+\{J^{i\,(1)},J^{j\,(1)}\},
   \end{gather}
where $g^{\mu\nu}=\mathrm{diag}(1,-1,-1,-1)$ is the metric tensor of Minkowski spacetime.   
The product $\bar{\tau}(p)\tau(p)=\tau(p)\bar{\tau}(p)$ reduces to
\begin{equation}
    \bar{\tau}(p)\tau(p)
    =(p^{\mu}p_{\mu})^2=p^4,
\end{equation}
for an arbitrary momentum $p^{\mu}$, and on shell to $\bar{\tau}(p)\tau(p)
    =m^4$.
With the aid of $\tau(p)$ and $\bar{\tau}(p)$, the left and right boost operators $B_{L}(\mathbf{p})$, $B_{R}(\mathbf{p})$ become
\begin{align}
    B_{L}(\mathbf{p})
    =\sqrt{\tau(p)}/m,\quad  B_{R}(\mathbf{p})
    =\sqrt{\bar{\tau}(p)}/m.
\end{align}
In terms of $\tau^{\mu\nu}$ and $\bar{\tau}^{\mu\nu}$ the Lorentz generators for the $(1,0)$ and $(0,1)$ irreps are, respectively 
\begin{align}
M_{L}^{\mu\nu}&=\frac{i}{6}\left( \tau^{\mu\rho}\bar{\tau}^{\nu}{}_{\rho}- \tau^{\nu\rho}\bar{\tau}^{\mu}{}_{\rho} \right),\\
M_{R}^{\mu\nu}&=\frac{i}{6}\left( \bar{\tau}^{\mu\rho}\tau^{\nu}{}_{\rho}- \bar{\tau}^{\nu\rho}\tau^{\mu}{}_{\rho} \right).
\end{align}
The Lorentz generators of the $(1,0)\oplus(0,1)$ representation in the chiral basis are given by 
\begin{equation}
M^{\mu\nu}=\begin{pmatrix}
    M_{L}^{\mu\nu} & 0\\
    0 & M_{R}^{\mu\nu}  
\end{pmatrix}.
\end{equation}
In this representation, one can define \emph{Parity} as the transformation $\Pi^{-1}\mathbf{A}\Pi=\mathbf{B}$, $\Pi^{-1}\mathbf{B}\Pi=\mathbf{A}$, and so it
exchanges $(a,b)\leftrightarrow (b,a)$ irreps.
In this basis, the boost operator for this representation is 
\begin{equation}\label{Boost1}
B(\mathbf{p})=\begin{pmatrix}
    B_{L}(\mathbf{p}) & 0\\
    0 & B_{R}(\mathbf{p})  
\end{pmatrix},
\end{equation}
while the parity operator reads
\begin{equation}
\Pi=\begin{pmatrix}
0 & 1\\
1 & 0
\end{pmatrix},
\end{equation}
and satisfies $\Pi=\Pi^{-1}=\Pi^\dagger=\Pi^T=\Pi^*$.
As shown in the following, this operator is a convenient tool in the $(1,0)\oplus(0,1)$ representation for the construction of the most general field that satisfies the KG equation.

\subsection{Parity Eigenvectors, Charge Conjugation and Chiral Operator}

In this subsection, we build explicitly the most general spin-1 field satisfying the KG equation and transforming under the $(1,0)\oplus(0,1)$ rep of the RLG. As a first step, we identify the eigenvectors of the parity operator $\Pi$, which are six-component column matrices, denoted here as $u^s_{\mathbf{0}}$ and $v^s_{\mathbf{0}}$
\begin{equation}
    \Pi u^s_{\mathbf{0}}=u^s_{\mathbf{0}},\qquad \Pi v^s_{\mathbf{0}}=-v^s_{\mathbf{0}},
\end{equation}
and they are given, in blocks, by
\begin{align}\label{ParEig}
    u_{\mathbf{0}}^{s}
    \equiv&
    m\begin{pmatrix}
        \xi^{s}
        \\
        \xi^{s}
    \end{pmatrix}
    &
    v_{\mathbf{0}}^{s}
    \equiv&
    m\begin{pmatrix}
        \chi^{s}
        \\
        -\chi^{s}
    \end{pmatrix},
\end{align}
where $\xi^{s}$, $s={1,0,-1}$ are the unit column vectors
\begin{align}
    \xi^{1}
    =&
    \begin{pmatrix}
        1\\0\\0
    \end{pmatrix},
    &
    \xi^{0}
    =&
    \begin{pmatrix}
        0\\1\\0
    \end{pmatrix},
    &
    \xi^{-1}
    =&
    \begin{pmatrix}
        0\\0\\1
    \end{pmatrix},
\end{align}
satisfying the orthonormality and completeness relations
\begin{align}
    \xi^{r\dagger}\xi^{s}
    =&
    \delta^{rs},
    &
    \sum_{s}\xi^{s}_{a}\xi^{s\dagger}_{b}
    =&
    \delta_{ab},
\end{align}
and $\chi^{s}=\epsilon\xi^{s\,*}$ are 
\begin{gather}
    \chi^{1}
    =
    \begin{pmatrix}
        0\\0\\1
    \end{pmatrix}
    =
    \xi^{-1},
    \quad
    \chi^{0}
    =
    \begin{pmatrix}
        0\\\text{-}1\\0
    \end{pmatrix}
    =
    -\xi^{0},\quad
    \chi^{-1}
    =
    \begin{pmatrix}
        1\\0\\0
    \end{pmatrix}
    =
    \xi^{1},
\end{gather}
or equivalently 
\begin{equation}\label{chixi}
\chi^s=(-1)^{s+1}\xi^{-s}.
\end{equation}

Boosting the parity operator, we obtain
\begin{equation}
\label{BoostingParity}
\begin{split}
    B(\mathbf{p}) \Pi B^{-1}(\mathbf{p})
    =&
    \frac{1}{m^{2}}
    S^{\mu\nu}
    p_{\mu}p_{\nu}
    \equiv
    \frac{1}{m^{2}}
    S(p),
\end{split}
\end{equation}
where we have defined the matrices
\begin{equation}
\label{SmunuCovariant}
    S^{\mu\nu}
    \equiv
    \begin{pmatrix}
        0 &\tau^{\mu\nu}\\
         \bar{\tau}^{\mu\nu} & 0
    \end{pmatrix}.   
\end{equation}
In terms of $S^{\mu\nu}$ the Lorentz generators for the $(1,0)\oplus(0,1)$ representation are
\begin{align}
M^{\mu\nu}&=\frac{i}{6}[S^{\mu\rho},S^{\nu}{}_{\rho}],
\end{align}
with \begin{gather}
M^{ij}\equiv\epsilon^{ijk}J^k,\qquad  M^{0i}=-M^{i0}\equiv K^i.
\end{gather}
Notice that $S^{00}=\Pi$.

In order to obtain the vectors $u_{\mathbf{p}}^{s}$ and $v_{\mathbf{p}}^{s}$ with arbitrary momentum $\mathbf{p}$ we have to apply the boost operator in Eq.~\eqref{Boost1} on the parity eigenvectors in Eq.~\eqref{ParEig}. The boosted parity eigenvectors $u_{\mathbf{p}}^{s}$ and $v_{\mathbf{p}}^{s}$ are then
\begin{gather}
    u_{\mathbf{p}}^{s}
    =
    B(\mathbf{p}) u_{\mathbf{0}}^{s}
    =
    m
    \begin{bmatrix}
        B_{L}(\mathbf{p}) \xi^{s} \\
        B_{R}(\mathbf{p}) \xi^{s}
    \end{bmatrix}
    =
\begin{bmatrix}
        \sqrt{\tau(p)} \xi^{s} \\
        \sqrt{\bar{\tau}(p) }\xi^{s}
    \end{bmatrix},
\\
    v_{\mathbf{p}}^{s}
    =
    B(\mathbf{p}) v_{\mathbf{0}}^{s}
    =
    m
    \begin{bmatrix}
        B_{L}(\mathbf{p}) \chi^{s} \\
        -B_{R}(\mathbf{p}) \chi^{s}
    \end{bmatrix}
    =
\begin{bmatrix}
        \sqrt{\tau(p)} \chi^{s} \\
        -\sqrt{\bar{\tau}(p)} \chi^{s}
    \end{bmatrix}
.
\end{gather}
Lorentz scalars can be built with the aid of the adjoints as follows
\begin{align}
    \overline{u}_{\mathbf{p}}^{s}=& u_{\mathbf{p}}^{s\,\dagger}S^{00}
    =
    \begin{bmatrix}
        \xi^{s\,\dagger}\sqrt{\bar{\tau}(p)}  &
        \xi^{s\,\dagger}\sqrt{\tau(p)}  
    \end{bmatrix} ,\\
    \overline{v}_{\mathbf{p}}^{s}=& v_{\mathbf{p}}^{s\,\dagger}S^{00}
    =
    \begin{bmatrix}
        -\chi^{s\,\dagger}\sqrt{\bar{\tau}(p) }&
       \chi^{s\,\dagger}\sqrt{ \tau(p) }
    \end{bmatrix}.
\end{align}

The charge-conjugate boosted parity eigenvectors $u_{\mathbf{p}}^{s\,c}$ and $v_{\mathbf{p}}^{s\,c}$ are defined as
\begin{align}\label{charcon}
    u_{\mathbf{p}}^{s\,c}
    \equiv &
    \mathcal{C}
    \overline{u}_{\mathbf{p}}^{s\,T},
    &
    v_{\mathbf{p}}^{s\,c}
    \equiv &
    \mathcal{C}\overline{u}_{\mathbf{p}}^{s\,T},
\end{align}
with 
\begin{align}
    \mathcal{C}=&
    \begin{pmatrix}
        \epsilon & 0 \\
         0 & \epsilon
    \end{pmatrix},
\end{align}
satisfying $\mathcal{C}=\mathcal{C}^{-1}=\mathcal{C}^\dagger=\mathcal{C}^{T}=\mathcal{C}^{*}$ and $[\mathcal{C},\Pi]=0$.
Explicitly
\begin{align}
    u_{\mathbf{p}}^{s\,c}= &
    m
    \begin{bmatrix}
       \epsilon B_{R}^{*}(\mathbf{p})\xi^{s\,*}
        \\
       \epsilon B_{L}^{*}(\mathbf{p})\xi^{s\,*}
    \end{bmatrix},
    &
    v_{\mathbf{p}}^{s\,c}= &
    m
    \begin{bmatrix}
       - \epsilon B_{R}^{*}(\mathbf{p})\chi^{s\,*}
        \\
       \epsilon B_{L}^{*}(\mathbf{p})\chi^{s\,*}
    \end{bmatrix}.
\end{align}
By combining Eq.~\eqref{conjeps} and Eq.~\eqref{BoostLR}, it can be shown that
\begin{align}
    \epsilon B^{*}_{R}(\mathbf{p})\epsilon=&B_{L}(\mathbf{p}),
    &
    \epsilon B^{*}_{L}(\mathbf{p})\epsilon=&B_{R}(\mathbf{p}),
\end{align}
and therefore, the charge-conjugate boosted parity eigenvectors become
\begin{align}
    u_{\mathbf{p}}^{s\,c}
    =& 
    m
    \begin{bmatrix}
        B_{L}(\mathbf{p})\chi^{s}
        \\
        B_{R}(\mathbf{p})\chi^{s}
    \end{bmatrix}
    =
\begin{bmatrix}
        \sqrt{\tau(p)} \chi^{s} \\
        \sqrt{\bar{\tau}(p)} \chi^{s}
    \end{bmatrix},
    \\
    v_{\mathbf{p}}^{s\,c}
    =&
    m
    \begin{bmatrix}
        -B_{L}(\mathbf{p})\xi^{s}
        \\
        B_{R}(\mathbf{p})\xi^{s}
    \end{bmatrix}
=
\begin{bmatrix}
       - \sqrt{\tau(p)} \xi^{s} \\
         \sqrt{\bar{\tau}(p)} \xi^{s}
    \end{bmatrix},
\end{align}
and their adjoints
\begin{align}
     \overline{u}_{\mathbf{p}}^{s\ c}=& u_{\mathbf{p}}^{s\,c\dagger}S^{00}
    =
    \begin{bmatrix}
        \chi^{s\,\dagger}\sqrt{\bar{\tau}(p)}  &
        \chi^{s\,\dagger}\sqrt{\tau(p)}  
    \end{bmatrix} ,\\
     \overline{v}_{\mathbf{p}}^{s\,c}= & v_{\mathbf{p}}^{s\,c\dagger}S^{00}
    =
    \begin{bmatrix}
        \xi^{s\,\dagger}\sqrt{\bar{\tau}(p) }&
      - \xi^{s\,\dagger}\sqrt{ \tau(p) }
    \end{bmatrix} .
\end{align}
Note that, in contrast with the Dirac theory, here the charge-conjugation operator does not change parity
\begin{equation}
u_{\mathbf{p}}^{s\,c}=(-1)^{s+1}u_{\mathbf{p}}^{-s},\qquad v_{\mathbf{p}}^{s\,c}=(-1)^{s}v_{\mathbf{p}}^{-s}.
\end{equation}

The chirality operator is defined as the Casimir operator
\begin{equation}
X=\frac{i}{8}\widetilde{M}^{\mu\nu}M_{\mu\nu},
\end{equation}
where $\widetilde{M}^{\mu\nu}=(1/2)\epsilon^{\mu\nu\rho\sigma}M_{\rho\sigma}$ are the dual Lorentz generators of the $(1,0)\oplus(0,1)$ representation, with $\epsilon^{\mu\nu\rho\sigma}$ being the totally antisymmetric Levi-Civita symbol in Minkowski space-time with $\epsilon^{0123}\equiv1$. In the chiral basis, the chirality operator becomes
\begin{equation}
X=\begin{pmatrix}
-1 & 0\\
0  & 1
\end{pmatrix}, 
\end{equation}
and it can be seen that this operator anti-commutes with the $S^{\mu\nu}$ matrices 
\begin{equation}
\{S^{\mu\nu},X\}=0.
\end{equation}
The vectors $u_{\mathbf{p}}^{s}$ and $u_{\mathbf{p}}^{s\,c}$ can be written in terms of $v_{\mathbf{p}}^{s}$ and $v_{\mathbf{p}}^{s\,c}$ using the chirality operator as follows
\begin{gather}
X u_{\mathbf{p}}^{s}
=v_{\mathbf{p}}^{s\,c}=(-1)^{s}v_{\mathbf{p}}^{-s},
\\
X u_{\mathbf{p}}^{s\,c}
=-v_{\mathbf{p}}^{s}=
(-1)^{s+1} v_{\mathbf{p}}^{-s\,c},
\\
\bar{u}_{\mathbf{p}}^{s} X=-\bar{v}_{\mathbf{p}}^{s\,c}=(-1)^{s+1} \bar{v}_{\mathbf{p}}^{-s}
\\
\bar{u}_{\mathbf{p}}^{s\,c} X=\bar{v}_{\mathbf{p}}^{s}=(-1)^{s}\bar{v}_{\mathbf{p}}^{-s\,c},
\end{gather}
and these vectors obey the following orthogonality relations
\begin{gather}
\bar{u}_{\mathbf{p}}^{r} u_{\mathbf{p}}^{s}=2m^2\delta^{rs},\quad
\bar{v}_{\mathbf{p}}^{r} v_{\mathbf{p}}^{s}=-2m^2\delta^{rs},
\\
\bar{u}_{\mathbf{p}}^{r\,c} u_{\mathbf{p}}^{s\,c}=2m^2\delta^{rs},
\quad
\bar{v}_{\mathbf{p}}^{r\,c} v_{\mathbf{p}}^{s\,c}=-2m^2\delta^{rs},\\
\bar{u}_{\mathbf{p}}^{r} v_{\mathbf{p}}^{s}=0,
\quad
\bar{v}_{\mathbf{p}}^{r} u_{\mathbf{p}}^{s}=0,\quad
\bar{u}_{\mathbf{p}}^{r\,c} v_{\mathbf{p}}^{s\,c}=0,
\quad
\bar{v}_{\mathbf{p}}^{r\,c} u_{\mathbf{p}}^{s\,c}=0,
\\
\bar{u}_{\mathbf{p}}^{r} v_{\mathbf{p}}^{s\,c}=0,
\quad
\bar{u}_{\mathbf{p}}^{r\,c} v_{\mathbf{p}}^{s}=0,
\quad
\bar{v}_{\mathbf{p}}^{r} u_{\mathbf{p}}^{s\,c}=0,
\quad
\bar{v}_{\mathbf{p}}^{r\,c} u_{\mathbf{p}}^{s}=0,
\\
\bar{u}_{\mathbf{p}}^{r\,c} u_{\mathbf{p}}^{s}=2m^2\chi^{r\,\dagger}\xi^{s},
\quad
\bar{u}_{\mathbf{p}}^{r} u_{\mathbf{p}}^{s\,c}=2m^2\xi^{r\,\dagger}\chi^{s},
\\
\bar{v}_{\mathbf{p}}^{r\,c} v_{\mathbf{p}}^{s}=-2m^2\xi^{r\,\dagger}\chi^{s},
\quad
\bar{v}_{\mathbf{p}}^{r} v_{\mathbf{p}}^{s\,c}=-2m^2\chi^{r\,\dagger}\xi^{s},
\end{gather}
together with the projections
\begin{gather}
\begin{split}
\sum_{s}
u_{\mathbf{p}a}^{s}
\bar{u}_{\mathbf{p}b}^{s}
=&
2
\left(
\Sigma^{\mu\nu}p_{\mu}p_{\nu}\right)_{ab}\equiv 2
\Sigma(p)_{ab},
\end{split}\\
\begin{split}
\sum_{s}
v_{\mathbf{p}a}^{s}
\bar{v}_{\mathbf{p}b}^{s}
=&
-2
\left(
\bar{\Sigma}^{\mu\nu}p_{\mu}p_{\nu}\right)_{ab}\equiv -2
\overline{\Sigma}(p)_{ab},
\end{split}
\\
\sum_{s}
u_{\mathbf{p}a}^{s\,c}
\bar{u}_{\mathbf{p}b}^{s\,c}=2
\Sigma(p)_{ab},
\\
\sum_{s}
v_{\mathbf{p}a}^{s\,c}
\bar{v}_{\mathbf{p}b}^{s\,c}=-2
\overline{\Sigma}(p)_{ab},
\end{gather}
where we have defined the operators~\cite{Napsuciale:2015kua}
\begin{equation}
\Sigma^{\mu\nu}=\frac{1}{2}(g^{\mu\nu}\mathbf{1}+S^{\mu\nu}),\qquad \overline{\Sigma}^{\mu\nu}=\frac{1}{2}(g^{\mu\nu}\mathbf{1}-S^{\mu\nu}).
\end{equation}
On shell, the boosted parity eigenvectors $u_{\mathbf{p}}^{s},v_{\mathbf{p}}^{s}$ and $u_{\mathbf{p}}^{s\,c},v_{\mathbf{p}}^{s\,c}$ satisfy the equations
\begin{align}
\label{sigmaeq1}
    \left[
    \Sigma(p)
    -m^{2}
    \right]
    u_{\mathbf{p}}^{s}
    =&0,
    &
    \left[
    \Sigma(p)
    -m^{2}
    \right]
    u_{\mathbf{p}}^{s\,c}
    =&0,
    \\
    \label{sigmaeq2}
    \left[
    \overline{\Sigma}(p)
    -m^{2}
    \right]
    v_{\mathbf{p}}^{s}
    =&0,
    &
    \left[
    \overline{\Sigma}(p)
    -m^{2}
    \right]
    v_{\mathbf{p}}^{s\,c}
    =&0.
\end{align}
An important observation is that, for an off shell momentum $p^\mu$, the product $[\Sigma(p)-m^2][\bar{\Sigma}(p)-m^2]=[\bar{\Sigma}(p)-m^2][\Sigma(p)-m^2]$ evaluates to
\begin{equation}
\begin{split}
[\Sigma(p)-m^2][\bar{\Sigma}(p)-m^2]
=-m^2(p^2-m^2)\mathbf{1},
\end{split}
\end{equation} 
thus, the boosted parity eigenvectors $u_{\mathbf{p}}^{s}$, $u_{\mathbf{p}}^{s\,c}$, $v_{\mathbf{p}}^{s}$ and $v_{\mathbf{p}}^{s\,c}$ are all solutions to the on-shell condition $(p^2-m^2)w_{\mathbf{p}}^{s}=0$, and therefore the most general field in the $(1,0)\oplus(0,1)$ representation of the RLG that satisfies the KG equation can be written as
\begin{equation}\label{field}\begin{split}
\psi(x)=&\int \frac{d^3\mathbf{p}}{(2\pi)^{3}2m\sqrt{\omega_{\mathbf{p}}}}\sum_s \bigg\{\left[u^{s}_{\mathbf{p}}a^{1 s}_{\mathbf{p}}+v^{s}_{\mathbf{p}}a^{2 s}_{\mathbf{p}}\right]e^{-ip\cdot x}\\&\qquad+\left[u^{s\,c}_{\mathbf{p}}b^{1 s\dagger}_{\mathbf{p}}+v^{s\, c}_{\mathbf{p}}b^{2 s\dagger}_{\mathbf{p}}\right]e^{ip\cdot x}\bigg\}.
\end{split}
\end{equation}

\section{Canonical Quantization}
\label{sec:quant}
\subsection{Pseudo-Hermitian Theory}
The field in Eq.~\eqref{field} can be successfully quantized as a pseudo-Hermitian quantum field theory, where the hermiticity of the Lagrangian is relaxed to the less stringent condition
\begin{equation}
\mathcal{L}^{\#}\equiv \eta^{-1}\mathcal{L}^\dagger\eta=\mathcal{L},
\label{pseudo-hermitian}
\end{equation}
for some operator $\eta$. 
The corresponding Lagrangian for our spin-1 SOPHY is given by
\begin{equation}
\mathcal{L}= \partial^{\mu}\widehat{\psi}\partial_\mu\psi-m^2\widehat{\psi}\psi ,
\label{Lag1}
\end{equation}
where $\widehat{\psi}=\eta^{-1}\bar\psi\eta$ is the redefined dual field that enforces the pseudo-Hermiticity of the quantum field theory~\cite{Sablevice:2023odu}, and is given by 
\begin{equation}\label{dual}
\begin{split}
\widehat{\psi}(x)
=&\int \frac{d^3\mathbf{p}}{(2\pi)^{3}2m\sqrt{\omega_{\mathbf{p}}}}\sum_s \bigg\{\left[\bar{u}^{s}_{\mathbf{p}}a^{1 s\dagger}_{\mathbf{p}}-\bar{v}^{s}_{\mathbf{p}}a^{2 s\dagger}_{\mathbf{p}}\right]e^{ip\cdot x}\\&\qquad+\left[\bar{u}^{s\,c}_{\mathbf{p}}b^{1 s}_{\mathbf{p}}-\bar{v}^{s\,c}_{\mathbf{p}}b^{2 s}_{\mathbf{p}}\right]e^{-ip\cdot x}\bigg\},
\end{split}
\end{equation}
the action of the operator $\eta $ is defined by the following relations
\begin{equation}
\begin{array}{cccc}
\eta^{-1} a^{j s}_{\mathbf{p}}\eta=(-1)^{j-1} a^{j s}_{\mathbf{p}},\qquad \eta^{-1} b^{j s\dagger}_{\mathbf{p}}\eta= (-1)^{j-1} b^{j s\dagger}_{\mathbf{p}},
\end{array}
\end{equation}
where $j=1,2$, and the explicit solution for this operator is found to be
\begin{equation}
\eta=\exp\left[i\pi\int\frac{d^3\mathbf{p}}{(2\pi)^{3}}\sum_s \left(a^{2 s\dagger}_{\mathbf{p}}a^{2 s}_{\mathbf{p}}+b^{2 s\dagger}_{\mathbf{p}}b^{2 s}_{\mathbf{p}}\right)\right],
\end{equation}
this operator satisfies $\eta=\eta^{-1}=\eta^\dagger$, meaning that it is simultaneously Hermitian and unitary, yielding $\eta^2=1$.

If the conjugate momenta $\pi_\psi=\dot{\widehat{\psi}}$ and $\pi_{\widehat{\psi}}=\dot{\psi}$ satisfy the equal-time canonical commutation relations
 \begin{equation}\label{canonicalquantization}
 \begin{split}
\left[\psi_{a}(\mathbf{x},t),\pi_\psi{}_{b}(\mathbf{x}',t)\right]&=\left[\widehat{\psi}_{a}(\mathbf{x},t),\pi_{\widehat{\psi}}{}_{b}(\mathbf{x}',t)\right]=i\delta_{ab}\delta^{(3)}(\mathbf{x}-\mathbf{x}'),
\end{split}
\end{equation}
then the momentum-space field operators satisfy the following commutation relations
\begin{equation}
\begin{split}
\left[a^{j s}_{\mathbf{p}},a^{k r\dagger}_{\mathbf{p}'}\right]&=(2\pi)^3\delta^{j k}\delta^{r s}\delta^{(3)}(\mathbf{p}-\mathbf{p}'),\\
\left[b^{j s}_{\mathbf{p}},b^{k r\dagger}_{\mathbf{p}'}\right]&=(2\pi)^3\delta^{j k}\delta^{r s}\delta^{(3)}(\mathbf{p}-\mathbf{p}').
\end{split}
\label{canonicalrelations}
\end{equation}
The resulting quantum field theory is causal, as can be verified explicitly 
\begin{equation}\label{causal}
\begin{split}
[\psi_a(x),\widehat{\psi}{}_{b}(x')]
&=\delta_{ab}\int \frac{d^3\mathbf{p}}{(2\pi)^{3}2\omega_{\mathbf{p}}} \bigg\{e^{-ip\cdot(x-x')}-e^{ip\cdot(x-x')}\bigg\}\\&=\delta_{ab}\Delta(x-x'),
\end{split}
\end{equation}
where $\Delta(x-x')$ is Schwinger's green function.
Similarly, we have $[\psi_a(x),\psi_{b}(x')]=0$ and $[\widehat{\psi}{}_a(x),\widehat{\psi}{}_{b}(x')]=0$, and the Feynman propagator can be straightforwardly found as 
\begin{equation}
\begin{split}
&[S_F(x-y)]_{\alpha\beta}=\bra{0}T[\psi_\alpha(x)\widehat{\psi}_\beta(y)]\ket{0}\\
&=\bra{0}[\theta(x^0-y^0)\psi_\alpha(x)\widehat{\psi}_\beta(y)+\theta(y^0-x^0)\widehat{\psi}_\beta(y)\psi_\alpha(x)]\ket{0}\\
&=\int\frac{d^4p}{(2\pi)^4}\left[\frac{i\delta_{\alpha\beta}}{p^2-m^2+i\epsilon}\right]e^{-i p\cdot(x-y)},
\end{split}
\end{equation}
as expected for a field that satisfies the KG equation.
\subsection{Noether charges}
The Hamiltonian and the momentum operator are given by
\begin{equation}
\begin{split}
H&=:\int d^3\mathbf{x}\bigg\{\dot{\widehat{\psi}}\dot{\psi}+\nabla\widehat\psi\cdot\nabla\psi+m^2\widehat\psi\psi\bigg\}:,\\
\mathbf{P}&=:-\int d^3\mathbf{x}\bigg\{\dot{\widehat{\psi}}\nabla\psi+\nabla\widehat{\psi}\dot{ \psi}\bigg\}:,
\end{split}
\end{equation}  
where $:{}:$ indicates normal ordering. Defining $P^\mu=(H,\mathbf{P})$, explicit calculation shows that
\begin{equation}
\begin{split}
     P^\mu&=\int  \frac{d^3\mathbf{p}}{(2\pi)^{3}}p^\mu\sum_{s,j}\left[a^{j s\dagger}_{\mathbf{p}}a^{j s}_{\mathbf{p}}+b^{j s\dagger}_{\mathbf{p}}b^{j s}_{\mathbf{p}}\right].
\end{split}    
\end{equation}
As can be seen from the above expression, the energy is bounded from below and, written in terms of the momentum-space operators, the Hamiltonian is in fact Hermitian.

Since the field in Eq.~\eqref{field} is complex, the theory has a global charge $\mathrm{U}(1)$ associated with the transformation
\begin{equation}
 \psi\rightarrow\psi'=e^{-i\alpha}\psi,\qquad  \widehat{\psi}\rightarrow\widehat{\psi}'=\widehat{\psi} e^{i\alpha},
\end{equation}
this conserved charge is given by
\begin{equation}
\begin{split}
Q=&:i\int d^3\mathbf{x}\bigg\{\widehat{\psi}\dot{\psi}-\dot{\widehat{\psi}}\psi
  \bigg\}:\\=&\int  \frac{d^3\mathbf{p}}{(2\pi)^{3}}\omega_{\mathbf{p}}\sum_{j,s}\left[a^{j s\dagger}_{\mathbf{p}}a^{j s}_{\mathbf{p}}-b^{j s\dagger}_{\mathbf{p}}b^{j s}_{\mathbf{p}}\right].
\end{split}    
\end{equation}
From the commutation relations of this operator with the creation and annihilation operators, one can conclude that $a^{j s \dagger}_{\mathbf{p}}$ and $b^{j s}_{\mathbf{p}}$ have charge $+1$, while $a^{j s }_{\mathbf{p}}$ and $b^{j s \dagger}_{\mathbf{p}}$ have charge $-1$.
The one-particle states can be labeled with this charge eigenvalue, and one can show that these states are twelve-fold degenerate
\begin{equation}
\begin{split}
H a^{j s\dagger}_{\mathbf{p}}\ket{0}&\propto H\ket{\mathbf{p},+,j,s}= \omega_{\mathbf{p}}\ket{\mathbf{p},+,j,s},\\
H b^{j s\dagger}_{\mathbf{p}}\ket{0}&\propto H\ket{\mathbf{p},-,j,s}= \omega_{\mathbf{p}}\ket{\mathbf{p},-,j,s}.
\end{split}
\end{equation}
Actually, the free theory has a larger symmetry that can be identified as follows: using the commutation relation $[\psi_{\alpha}(x),\widehat{\psi}_{\beta}(x)]=0$, following from Eq.~\eqref{causal}, the Lagrangian Eq.~\eqref{Lag1} can be recast in the form
\begin{equation}
\mathcal{L}=\frac{1}{2}\partial^\mu\Psi^T\Xi\partial_\mu\Psi-\frac{m^2}{2}\Psi^T\Xi\Psi,\label{Lag2}
\end{equation}
where $\Psi$ is the column vector
\begin{equation}
\Psi(x)=\begin{pmatrix}
\widehat{\psi}^T(x)\\
\psi(x)
\end{pmatrix},
\end{equation}
and $\Xi$ is a $12 \times 12$ matrix, written in $6 \times 6$ blocks as
\begin{equation}
\Xi = \begin{pmatrix}
0_{6 \times 6} & 1_{6 \times 6} \\
1_{6 \times 6} & 0_{6 \times 6}
\end{pmatrix}=\sigma^1\otimes 1_{6 \times 6}.
\end{equation}
The Lagrangian in Eq.~\eqref{Lag2} is invariant under global transformations of the form $\Psi \to \Psi' = M \Psi$ with $M^T \Xi M = \Xi$. This group turns out to be isomorphic to the group $O(12,\mathbb{C})$ which has 66 independent generators $X$, satisfying $X^T\Xi+\Xi X=0$. A convenient basis for these generators is 
\begin{equation}
1_{2 \times 2}\otimes A_1,\qquad \sigma^1\otimes A_2,\qquad \sigma^2\otimes A_3,\qquad \sigma^3\otimes S,
\end{equation}
where $A_i, S$, $i,=1,2,3$ are arbitrary antisymmetric, and symmetric $6 \times 6$ matrices, respectively. 

It is important to remark that the theory is, by construction, a SOPHY for spin-$1$ fields. It is possible to explicitly show that the field indeed contains particles of spin $1$
by evaluating the action of the spin operator
\begin{equation}
\begin{split}
\mathbf{S}=:-i\int d^3\mathbf{x}\left\{ \dot{\widehat{\psi}}\mathbf{J}\psi-\widehat{\psi}\mathbf{J}\dot{\psi}\right\}:,
\end{split}
\end{equation}
with
\begin{equation}
    \mathbf{J}=\begin{pmatrix}
    \mathbf{J}^{(1)}&0\\
    0& \mathbf{J}^{(1)}
    \end{pmatrix},
\end{equation}
on one-particle states with zero momentum in the chiral representation. The resulting relations are
\begin{equation}
S^3a^{j s\dagger}_{\mathbf{0}}\ket{0}= s a^{j s\dagger}_{\mathbf{0}}\ket{0},\quad S^3b^{j s\dagger}_{\mathbf{0}}\ket{0}= s b^{j s\dagger}_{\mathbf{0}}\ket{0}.
\end{equation}

\subsection{Discrete symmetries}
The transformation properties of the field and its dual in Eqs.(\ref{field},\ref{dual}) under the full Lorentz Group $O(1,3)$ are determined by the study of their behavior under the action of Parity $P$, Time-reversal $T$ and Charge conjugation $C$.
Parity reverses the three-momentum and leaves the spin invariant. Thus, its action on creation and annihilation operators is
\begin{align}
    P^{-1}a^{j s\dagger}_{\mathbf{p}}P&=(-1)^{j-1}a^{j s\dagger}_{-\mathbf{p}},\\
    P^{-1}b^{j s\dagger}_{\mathbf{p}}P&=(-1)^{j-1}b^{j s\dagger}_{-\mathbf{p}},
\end{align}
where the phases have been chosen such that our fields belong to the usual Wigner class discussed by Lee and Wick in~\cite{Lee:1966ik}, by noticing that the inverse boost operator is
\begin{equation}
\begin{split}
    B(-\mathbf{p})
    =&S^{00}B(\mathbf{p})S^{00}\\=&  
    \begin{pmatrix}
        B_{R}(\mathbf{p}) & 0
        \\
        0 & B_{L}(\mathbf{p})
    \end{pmatrix}
          =
    \begin{pmatrix}
        \sqrt{\overline{\tau}(p)} & 0
        \\
        0 & \sqrt{\tau(p)}
    \end{pmatrix},
\end{split}
\end{equation}
and therefore, the following relations hold
\begin{gather}
u^{s}_{-\mathbf{p}}=S^{00}u^{s}_{\mathbf{p}},\quad
u^{s\,c}_{-\mathbf{p}}=S^{00}u^{s\,c}_{\mathbf{p}},\\
v^{s}_{-\mathbf{p}}=-S^{00}v^{s}_{\mathbf{p}},\quad
v^{s\,c}_{-\mathbf{p}}=-S^{00}v^{s\,c}_{\mathbf{p}},
\end{gather}
yielding
\begin{align}
&P^{-1}\psi(x)P
=S^{00}\psi(\mathcal{P} x),
\end{align}
with $\mathcal{P}=\mathrm{diag}(1,-1,-1,-1)$.
Similarly, the parity transformation for the dual field is given by
\begin{align}
&P^{-1}\widehat{\psi}(x)P=\widehat{\psi}(\mathcal{P}x)S^{00}.
\end{align}

Another feature is that charge conjugation interchanges particles and antiparticles as can be seen as follows
\begin{align}
    C^{-1}a^{j s\dagger}_{\mathbf{p}}C&=(-1)^{j-1}b^{j s\dagger}_{\mathbf{p}},\\
    C^{-1}b^{j s\dagger}_{\mathbf{p}}C&=(-1)^{j-1}a^{j s\dagger}_{\mathbf{p}}.
\end{align}
Using the definition of Charge conjugation in Eq.~\eqref{charcon}, the corresponding transformation of the field and its dual reads
\begin{align}
&C^{-1}\psi(x)C=\mathcal{C}\widehat{\psi}^T(x),&C^{-1}\widehat{\psi}C=\psi^T(x)\mathcal{C}.
\end{align}
Notice that in the chiral basis we have $\mathcal{C}=-S^{00}S^{22}=-S^{22}S^{00}=S^{11}S^{33}$. 

Finally, time-reversal changes the direction of both momentum and spin 
\begin{align}
    T^{-1}a^{j s\dagger}_{\mathbf{p}}T&=(-1)^{s+1}a^{j -s\dagger}_{-\mathbf{p}},\\
    T^{-1}b^{j s\dagger}_{\mathbf{p}}T&=(-1)^{s+1}b^{j -s\dagger}_{-\mathbf{p}}.
\end{align}
Using the identities
\begin{align}
    u^{-s*}_{-\mathbf{p}}&=(-1)^{s+1}\mathcal{C}u^{s}_{\mathbf{p}},\\
    v^{-s*}_{-\mathbf{p}}&=(-1)^{s+1}\mathcal{C}v^{s}_{\mathbf{p}},\\
    u^{-s\, c*}_{-\mathbf{p}}&=(-1)^{s+1}\mathcal{C}u^{s\,c}_{\mathbf{p}},\\
    v^{-s\,c*}_{-\mathbf{p}}&=(-1)^{s+1}\mathcal{C}v^{s\,c}_{\mathbf{p}},
\end{align}
that follow from $B^*(-\mathbf{p})=\mathcal{C} B(\mathbf{p})\mathcal{C} $, and Eq.~\eqref{chixi},
the time-reversal transformations for the field and its dual become
\begin{align}
&T^{-1}\psi(x)T=\mathcal{C}\psi(\mathcal{T}x),&T^{-1}\widehat{\psi}(x)T=\widehat{\psi}(\mathcal{T}x)\mathcal{C}.
\end{align}
where we have defined
$\mathcal{T}=\mathrm{diag}(-1,1,1,1)$. Thus, our second-order theory is invariant under Parity ($\mathrm{P}$), Charge conjugation ($\mathrm{C}$) and time reversal ($\mathrm{T}$), and therefore under $\mathrm{CPT}$.

\subsection{Interactions}
The simplest $\mathrm{C}$, $\mathrm{P}$ and $\mathrm{T}$ invariant pseudo-Hermitian interactions that can be introduced in this framework are self-interactions represented by a dimension-four renormalizable operator
\begin{equation}
\begin{split}
\mathcal{L}_{\text{int}}=&   \frac{\lambda_{1}}{2}\left(\widehat{\psi}\psi\right)^{2}
+ \frac{\lambda_{2}}{2}\left( \widehat{\psi}X\psi\right)  \left(\widehat{\psi}X\psi\right)\\
&+ \frac{\lambda_{3}}{2}\left( \widehat{\psi}M^{\mu\nu}\psi\right)  \left(\widehat{\psi}M_{\mu\nu}\psi\right)\\
&+ \frac{\lambda_{4}}{2}\left( \widehat{\psi}S^{\mu\nu}\psi\right)  \left(\widehat{\psi}S_{\mu\nu}\psi\right).
\label{Lag3}
\end{split}
\end{equation}
The renormalization of these interactions was studied in~\cite{Rivero-Acosta:2020hvy}, in the context of the naïve Hermitian second-order theory, with the six complex components of the field packed into a complex second-rank antisymmetric tensor. Nevertheless, all the results contained in~\cite{Rivero-Acosta:2020hvy} can be straightforwardly generalized to the SOPHY version of the theory presented here, showing that this class of theories has a rich set of renormalization group equations and constitute a valuable theoretical playground for renormalization studies. Recently, in \cite{Leng:2025qlu}, the unitary evolution of pseudo-Hermitian quantum field theories was investigated, setting the foundations for a consistent formulation of scattering processes. This work will be crucial in the analysis of the self interactions in Eq.(\ref{Lag3}), to be presented in a forthcoming paper.

Moreover, since $\psi$ has mass dimension one, if it transforms as a singlet under the SM symmetries, it cannot decay into SM fields. Therefore, the least massive pseudo-Hermitian particle is automatically stable and could play the role of a WIMP dark matter candidate. The natural interaction of this field with the SM particles is through a Higgs portal of the form
\begin{equation}
\mathcal{L}_{\psi H}=   \lambda_{H}\left(\widehat{\psi}\psi\right)H^\dagger H\label{LagHiggs}.
\end{equation}
In this case the dark matter phenomenology is determined by the mass of the field $m$ and its coupling $\lambda_{H}$ in a similar way as for complex scalar dark matter~\cite{Feng:2014vea,Arcadi:2017kky}. The same interaction contributes to the running of the Higgs mass as an extra set of scalars does, and since the sign of $\lambda_H$ is undetermined, we can choose it in such a way as to cancel the top-quark contribution. This phenomenology and further consequences will be the subject of a forthcoming article.

\section{Conclusions}
\label{sec:conc}

In this work, we have studied the quantization of a massive field transforming in the $(1,0)\oplus(0,1)$ representation of the RLG that satisfies the KG equation. We have shown that the canonical quantization is rendered consistent by an adequate choice of the field dual and the introduction of a unitary and Hermitian operator $\eta$ that relaxes the Hermiticity condition into a less stringent pseudo-Hermitian one, dictated by Eq.~\eqref{pseudo-hermitian}. The resulting SOPHY is causal, and the Hamiltonian is bounded from below with real eigenvalues. For consistency, we have verified that the corresponding fields have spin $1$, and we have identified the global and discrete symmetries of the theory, which turns out to be invariant under $C$, $P$ and $T$. Finally, we have sketched the plausible interactions, including a rich class of renormalizable quartic self-interactions, and the identification of this field as a WIMP Dark Matter candidate through a Higgs portal. We conclude this paper by pointing out that the generalization of the present model to massive fields transforming in the $(j,0)\oplus(0,j)$ representation of the RLG is a straightforward exercise, and therefore this class of SOPHIES can shed valuable light on the study of consistent higher-spin theories.

\section*{Acknowledgements}
This work was supported  by the Secretar\'ia de Ciencia, Humanidades, tecnolog\'ia e Innovacion  (SECIHTI) through the project IxM 749. A. de la C. acknowledges SECIHTI national scholarships. I. D-S. was supported by the SECIHITI program ``Estancias Posdoctorales por M\' exico''.




\bibliographystyle{elsarticle-harv} 

\begin{thebibliography}{00}

\bibitem{LeClair:2007iy}
A.~LeClair and M.~Neubert,
JHEP \textbf{10}, 027 (2007)
doi:10.1088/1126-6708/2007/10/027
[arXiv:0705.4657 [hep-th]].


\bibitem{Mostafazadeh:2001jk}
A.~Mostafazadeh,
J. Math. Phys. \textbf{43}, 205-214 (2002)
doi:10.1063/1.1418246
[arXiv:math-ph/0107001 [math-ph]].

\bibitem{Mostafazadeh:2008pw}
A.~Mostafazadeh,
Int. J. Geom. Meth. Mod. Phys. \textbf{07}, 1191-1306 (2010)
doi:10.1142/S0219887810004816
[arXiv:0810.5643 [quant-ph]].

\bibitem{Bender:1998ke}
C.~M.~Bender and S.~Boettcher,
Phys. Rev. Lett. \textbf{80}, 5243-5246 (1998)
doi:10.1103/PhysRevLett.80.5243
[arXiv:physics/9712001 [physics]].

\bibitem{Christodoulides:2018arz}
D.~Christodoulides and J.~Yang,
Springer Tracts Mod. Phys. \textbf{280}, pp.1-579 (2018)
Springer, 2018,
doi:10.1007/978-981-13-1247-2

\bibitem{Ferro-Hernandez:2023ymz}
R.~Ferro-Hern\'andez, J.~Olmos, E.~Peinado and C.~A.~Vaquera-Araujo,
Phys. Rev. D \textbf{109} (2024) no.8, 085003
doi:10.1103/PhysRevD.109.085003
[arXiv:2311.06466 [hep-th]].

\bibitem{Ahluwalia:2023slc}
D.~V.~Ahluwalia, G.~B.~de Gracia, J.~M.~H.~da Silva, C.~Y.~Lee and B.~M.~Pimentel,
JHEP \textbf{04} (2024), 075
doi:10.1007/JHEP04(2024)075
[arXiv:2312.17038 [hep-th]].

\bibitem{deGracia:2024umr}
G.~B.~de Gracia, R.~da Rocha, R.~J.~Bueno Rogerio and C.~Y.~Lee,
Phys. Dark Univ. \textbf{47}, 101774 (2025)
doi:10.1016/j.dark.2024.101774
[arXiv:2407.00126 [hep-ph]].

\bibitem{deGracia:2025omq}
G.~B.~de Gracia and R.~J.~B.~Rogerio,
[arXiv:2511.13951 [gr-qc]].

\bibitem{Rangel-Pantoja:2025mhg}
A.~d.~Rangel-Pantoja, I.~D{\'\i}az-Salda{\~n}a and C.~A.~Vaquera-Araujo,
Universe \textbf{11}, no.12, 400 (2025)
doi:10.3390/universe11120400
[arXiv:2505.21708 [hep-th]].


\bibitem{Ahluwalia:2022zrm}
D.~V.~Ahluwalia and C.~Y.~Lee,
EPL \textbf{140} (2022) no.2, 24001
[erratum: EPL \textbf{140} (2022) no.6, 69901]
doi:10.1209/0295-5075/ac97bd
[arXiv:2212.09457 [hep-th]].

\bibitem{Bai:2025pvb}
Y.~Bai, C.~Y.~Lee, R.~Leng and S.~Zhou,
[arXiv:2510.22952 [hep-th]].


\bibitem{Englert:1964et}
F.~Englert and R.~Brout,
Phys. Rev. Lett. \textbf{13}, 321-323 (1964)
doi:10.1103/PhysRevLett.13.321

\bibitem{Higgs:1964pj}
P.~W.~Higgs,
Phys. Rev. Lett. \textbf{13}, 508-509 (1964)
doi:10.1103/PhysRevLett.13.508

\bibitem{Guralnik:1964eu}
G.~S.~Guralnik, C.~R.~Hagen and T.~W.~B.~Kibble,
Phys. Rev. Lett. \textbf{13}, 585-587 (1964)
doi:10.1103/PhysRevLett.13.585

\bibitem{Ruegg:2003ps}
H.~Ruegg and M.~Ruiz-Altaba,
Int. J. Mod. Phys. A \textbf{19}, 3265-3348 (2004)
doi:10.1142/S0217751X04019755
[arXiv:hep-th/0304245 [hep-th]].


\bibitem{Napsuciale:2022usn}
M.~Napsuciale,
Phys. Scripta \textbf{98} (2023) no.9, 095305
doi:10.1088/1402-4896/acecb5
[arXiv:2212.01153 [hep-ph]].


\bibitem{Langacker:2017uah}
P.~Langacker,
Taylor \& Francis, 2017,
ISBN 978-1-4987-6322-6, 978-1-4987-6321-9, 978-0-367-57344-7, 978-1-315-17062-6
doi:10.1201/b22175

\bibitem{Napsuciale:2015kua}
M.~Napsuciale, S.~Rodr\'\i{}guez, R.~Ferro-Hern\'andez and S.~G\'omez-\'Avila,
Phys. Rev. D \textbf{93}, no.7, 076003 (2016)
doi:10.1103/PhysRevD.93.076003
[arXiv:1509.07938 [hep-ph]].

\bibitem{Sablevice:2023odu}
E.~Sablevice and P.~Millington,
Phys. Rev. D \textbf{109} (2024) no.6, 6
doi:10.1103/PhysRevD.109.065012
[arXiv:2307.16805 [hep-th]].

\bibitem{Lee:1966ik}
T.~D.~Lee and G.~C.~Wick,
Phys. Rev. \textbf{148} (1966), 1385-1404
doi:10.1103/PhysRev.148.1385

\bibitem{Rivero-Acosta:2020hvy}
A.~Rivero-Acosta and C.~A.~Vaquera-Araujo,
Eur. Phys. J. C \textbf{80} (2020) no.7, 618
doi:10.1140/epjc/s10052-020-8190-5
[arXiv:2003.08320 [hep-ph]].

\bibitem{Leng:2025qlu}
R.~Leng, C.~Y.~Lee and S.~Zhou,
[arXiv:2510.27404 [hep-th]].

\bibitem{Feng:2014vea}
L.~Feng, S.~Profumo and L.~Ubaldi,
JHEP \textbf{03}, 045 (2015)
doi:10.1007/JHEP03(2015)045
[arXiv:1412.1105 [hep-ph]].

\bibitem{Arcadi:2017kky}
G.~Arcadi, M.~Dutra, P.~Ghosh, M.~Lindner, Y.~Mambrini, M.~Pierre, S.~Profumo and F.~S.~Queiroz,
Eur. Phys. J. C \textbf{78}, no.3, 203 (2018)
doi:10.1140/epjc/s10052-018-5662-y
[arXiv:1703.07364 [hep-ph]].




\end{thebibliography}


\end{document}